\documentstyle[prd,aps]{revtex}

\begin{document}

\draft

\title{The spacetime attribute of matter}

\author{Borge Nodland}

\address{Department of Physics and Astronomy,
University of Rochester,
Rochester, New York 14627}

\author{We propose that spacetime is fundamentally a property of 
matter, inseparable from it. This leads us to suggest that all 
properties of matter must be elevated to the same status as that of 
spacetime in quantum field theories of matter. We suggest a specific 
method for extending field theories to accomodate this, and point 
out how this leads to the evolution of fields through channels other 
than the spacetime channel.}

\address{PACS numbers: 11.10.-z, 03.65.Bz, 03.65.-w}

\date{December, 1997}

\maketitle

\section{Introduction} 

In general, an interacting quantum mechanical system is described by
an operator equation for the field amplitude $f$ of the system. This
field equation is of the form

\begin{equation}
\widehat{O}(\frac{\partial}{\partial x}, x, a_i) f(x) 
= I_{\text{int}}(x)
\label{eq1}
\end{equation}
(throughout this paper, we use an overhat to denote operators).
The spacetime-dependent term $I_{\text{int}}(x)$ in (\ref{eq1}) is
given by
\begin{equation}
I_{\text{int}}(x)= 
\frac{\partial {\cal L}_{\text{int}}}{\partial f},
\label{eq4}
\end{equation}
and arises from the Euler-Lagrange equation

\begin{equation}
\frac{\partial {\cal L}}{\partial f}
- \partial_{\mu} 
[\frac{\partial {\cal L}}{\partial (\partial_\mu f)}] = 0
\label{eq5}
\end{equation}
when interactions are present in the form of an interaction term ${\cal
L}_{\text{int}}$ in the Lagrangian density $\cal{L}$ \cite{ryd}.

We note that spacetime $x = x^\mu$ typically enter as a derivative
$\frac{\partial}{\partial x}$ (and also possibly as a parameter) in the
operator $\widehat{O}$ in (\ref{eq1}), while other properties, or
attributes, $a_i$, such as mass, spin, etc., generally enter only as
parameters, not derivatives (an example is the Dirac equation).
Spacetime, therefore, assumes a special role in (\ref{eq1}).

However, one may take the view that the spacetime location $x$ of a
particle represented by a field $f$ is just one of many attributes the
field can have. A separation of spacetime from objects or fields is
artificial, since space really has no meaning without physical objects
in it, and time has no meaning unless a physical change occurs in an
object. From these considerations, we propose that spacetime $x$ is
a regular property of matter, and we take this
as an indication that other properties of matter should play a role of
the same significance as that of spacetime in quantum field theories.
In other words, spacetime exemplifies how other attributes of matter
should be viewed in field theories.

A simple, natural approach toward the attainment of a satisfactory
quantum theory of gravitation is to regard spacetime $x$ as a discrete,
quantized entity (see for example Ref. \cite{hof} and references
therein). This accords with the quantal nature of other particle
properties like spin, mass, etc, and supports the notion that spacetime
is a property of matter.

An attribute of a field can exist only if the field exists. For
example, since the collection of all measurable values for the
spacetime location of a particle represented by the field $f$ is just
spacetime itself, spacetime exists only becuse objects, or fields,
exist. The same reasoning can be applied to other field attributes. A
field may thus be considered to carry with it several attributive
spaces, such as spacetime, a mass--space, a spin--space, etc.

The presence of spacetime derivatives in quantum field theory (i.e. in
Lagrangian densities, and consequently in generating functionals, field
equations, etc.) leads to an evolution of a field amplitude in space
and time via interactions with other fields. Interactions between
fields of specific masses, spins, and spatiotemporal location
probabilities result in new fields with new values for these
properties. Commonly, such interactions are thought to proceed through
space and time. However, if one considers spacetime to be one of many
attributive spaces carried by a field, one may consider interactions to
go through alternative channels. A specific such channel would then be
an attributive space existing by virtue of characterizing a field. The
totality of measurable values of the specific attribute of the field is
what constitutes the space or channel.  Examples of such alternative
channels are a mass (or energy) channel, and a spin channel.

\section{General Approach}

In view of the discussion in the introduction, we now extend the
interpretation of the field $f$ to that of a function $\widetilde{f}$
of several attributes (in this paper, we use a tilde throughout to
denote extensions of operators and fields). This means that
$\widetilde{f}$ is an attributive field defined on a multiattributive
space ${\cal S}$, containing as subspaces the attributive spaces
mentioned in the introduction (among others).  $\widetilde{f}$ is now a
multiparticle field that contains information on the evolution (via
interactions) of particles into new particles characterized by new
attributive values for their spacetime location, mass, spin, etc.

With this extension of $f$ in (\ref{eq1}), the operator $\widehat{O}$
in (\ref{eq1}) must similarly be extended to an operator
$\widetilde{\widehat{O}}$ that takes into account the other attributes
$f$ has in the same manner that the spacetime attribute $x$ is
normally taken into account in (\ref{eq1}). This implies that an
attribute $a_i$ should not merely be a parameter in (\ref{eq1}), but
should also enter as a derivative $\frac{\partial}{\partial a_i}$.

In view of the above, we therefore extend (\ref{eq1}) to the general
form 

\begin{equation}
\widetilde{\widehat{O}}(\frac{\partial}{\partial a_1}, 
\frac{\partial}{\partial a_2}, \cdots \;, a_1, a_2, \cdots \;) 
\widetilde{f}(a_1, a_2, \cdots \;) 
= 0,
\label{eq6}
\end{equation}
where $a_i \; (i =, 1, 2, \cdots \;)$ are the attributes of
$\widetilde{f}$, including the spacetime attribute $x$.

Since $\widetilde{f}$ is inherently a field of multiple, and thus
interacting, particles, the operator $\widetilde{\widehat{O}}$ is
assumed to account for all interactions. This obviates the need for a
separate interaction term [as in (\ref{eq1})] on the right hand side of
(\ref{eq6}). One may consider the field equation (\ref{eq1}) to be a
projection onto the spacetime subspace of ${\cal S}$ of an underlying
field equation (\ref{eq6}), where in this projection only a single
attributive derivative, that is, the spacetime derivative
$\frac{\partial}{\partial a_1} =
\frac{\partial}{\partial x} =
\frac{\partial}{\partial x^\mu} = 
\partial_\mu$ is retained.

In view of the discrete nature of an attribute $a_i$ (including $x$, as
explained above), the attributive space ${\cal S}$ is strictly speaking
a discrete space over which a difference operation (or possibly some
alternative operation) on $\widetilde{f}$ would be more appropriate
than differentiation. However, for simplicity, we still retain the
notation in terms of differentiation as in (\ref{eq1}), with the
understanding that it should be re--interpreted in terms of a
difference operation, consistent with the quantal nature of ${\cal
S}$. We also assume that this operation is linear.

In what follows, we want to utilize specific, established expressions
for the operator $\widehat{O}(\frac{\partial}{\partial x}, x, b)$ in
(\ref{eq1}) to obtain information on the
nature of an attribute $b$ that appears as a derivative in (\ref{eq6}),
while only entering as a parameter in (\ref{eq1}) [i.e. $b \neq x$].
One way to do this is to apply a differential operator
$\frac{\partial}{\partial b}$ to (\ref{eq1}). Thus, if (\ref{eq1})
[with $a_i$ replaced by $b$, and $f(x)$ replaced by $\widetilde{f}(x,
b)$] holds, then

\begin{equation}
\frac{\partial}{\partial b}
[\widehat{O}(\frac{\partial}{\partial x}, x, b) 
\widetilde{f}(x, b)] 
= 0
\label{eq7}
\end{equation}
must hold. It is seen that (\ref{eq7}) is a two-attribute version
of (\ref{eq6}),
\begin{equation}
\widetilde{\widehat{O}}(\frac{\partial}{\partial x}, 
\frac{\partial}{\partial b}, x, b) 
\widetilde{f}(x, b) = 0,
\label{eq8}
\end{equation}
if we extend the interpretation of the field $f(x)$ in (\ref{eq1})
to that of an attributive field $\widetilde{f}(x, b)$ over $x$ and
$b$.

\section{Application to the Dirac Equation}

As a specific application of the above discussion, we focus on
the Dirac equation

\begin{equation}
(i \gamma^\mu \partial_\mu - m)\psi(x) 
= I_{\text{int}}(x),
\label{eq8.5}
\end{equation}
where $\gamma^\mu \;\; (\mu=0,1,2,3)$ are the Dirac gamma matrices, $m$ the
mass of $\psi$, and $\partial_\mu=\frac{\partial}{\partial x}$ the
spacetime derivative operator. From (\ref{eq8.5}), we see that examples
of the non--spacetime attributes $a_i$ in (\ref{eq1}) are quantities like
$m$ (representing mass) and $\gamma^\mu$ (representing spin).

According to the discussion in the previous section, we now extend the interpretation of the Dirac field $\psi(x)$ in (\ref{eq8.5}) to
that of a field $\widetilde{\psi}(x, m)$ over spacetime $x$ and
mass $m$. We next differentiate (\ref{eq8.5}) [with $\psi(x)$
replaced by $\widetilde{\psi}(x, m)$] with respect to mass $m$,
corresponding to the operation in (\ref{eq7}), with the attribute $b$
of the general field $f$ being the mass attribute $m$ of the extended
Dirac field $\widetilde{\psi}$,

\begin{equation}
\frac{\partial}{\partial m}
(i \gamma^\mu \partial_\mu - m)\widetilde{\psi}(x, m) 
= \frac{\partial}{\partial m} I_{\text{int}}(x).
\label{eq9}
\end{equation}
This results in the equation

\begin{equation}
(i \gamma^\mu \partial_\mu - m)
\frac{\partial \widetilde{\psi}}{\partial m} 
- \widetilde{\psi} = 0.
\label{eq10}
\end{equation}

To solve this equation, we note that the Dirac operator

\begin{equation}
\widehat{D}(x, m) = (i \gamma^\mu \partial_\mu - m)
\label{eq11}
\end{equation}
has the inverse (i.e. the corresponding Dirac propagator)

\begin{equation}
S(x, m) = (i \gamma^\mu \partial_\mu + m) \Delta_F(x,m),
\label{eq12}
\end{equation}
since

\begin{eqnarray}
& &\widehat{D}(x, m) S(x,m) \nonumber \\
&=& (i \gamma^\mu \partial_\mu + m) 
(i \gamma^\nu \partial_\nu + m) \Delta_F(x, m) \nonumber \\
&=& (-\Box - m^2) \Delta_F(x,m) = \delta^4(x).
\label{eq12.3}
\end{eqnarray}
Here $\Delta_F(x, m)$ is the Feynman propagator, $\Box$ the contraction
$\partial^\mu \partial_\mu$, and $\delta(x)$ the four--dimensional
delta function. (\ref{eq12.3}) follows from the symmetry property
$\partial_\mu \partial_\nu = \partial_\nu \partial_\mu$ of the
derivative operator $\partial_\mu \partial_\nu$, the anticommutator
property $\{\gamma^\mu, \gamma^\nu\} = 2 g^{\mu\nu}$ [where
$g^{\mu\nu}$ is the metric of signature (1, -1, -1, -1)] of the gamma
matrices $\gamma^\mu$, and the fact that the Feynman propagator is the
inverse of the Klein--Gordon operator $\widehat{K}(x, m)$,

\begin{eqnarray}
\widehat{K}(x, m) \Delta_F(x, m) 
&=& (\Box + m^2) \Delta_F(x,m)\nonumber \\
&=& -\delta^4(x).
\label{eq12.5}
\end{eqnarray}

In accord with the general discussion in the previous section, we have
included here the specific $m$--dependence in all operators and
propagators. As with the general operator $\widehat{O}$, we assume that
its propagator inverse $\widehat{O}^{-1}$ has an extension
$\widetilde{\widehat{O}}^{-1}$ that is generally different from its
spacetime projection (i.e.  ${\widetilde{\widehat{O}}}^{-1} \neq
\widehat{O}^{-1}$).

Multiplication of $S(x - x_1, m)$ on both sides of (\ref{eq10}), and
integrating with respect to spacetime $x_1$ yields, with the use of
(\ref{eq12.3}),

\begin{equation}
\frac{\partial \widetilde{\psi}(x, m)}{\partial m} 
= \int S(x - x_1, m) \widetilde{\psi}(x_1, m) \, d^4x_1.
\label{eq12.6}
\end{equation}
We see that the integro-differential equation (\ref{eq12.6}) indicates
that $\widetilde{\psi}$ undergoes an interaction--driven evolution
through a mass channel, as discussed in the introduction. Consequently,
we define a mass evolution operator $\widehat{{\cal M}}(x, m, m_0)$ by

\begin{equation}
\widetilde{\psi}(x, m) 
= \widehat{{\cal M}}(x, m, m_0) \widetilde{\psi}(x, m_0),
\label{eq13}
\end{equation}
where $\widetilde{\psi}(x, m_0)$ is an extended Dirac field
characterizing a system with a fixed mass $m_0$. Clearly, we have

\begin{equation}
\widehat{{\cal M}}(x, m_0, m_0) = 1.
\label{eq14}
\end{equation}

In view of the definition (\ref{eq13}) for $\widehat{{\cal M}}$,
(\ref{eq12.6}) is equivalent to

\begin{equation}
\frac{\partial \widehat{{\cal M}}(x, m, m_0)}{\partial m} 
= \int S(x - x_1, m) \widehat{{\cal M}}(x_1, m, m_0) \, d^4x_1.
\label{eq15}
\end{equation}

But the differential equation (\ref{eq15}) and the boundary
condition (\ref{eq14}) can be combined to give the single
integral equation [with $x$ replaced by $x_0$]

\begin{eqnarray}
& & \widehat{{\cal M}}(x_0, m, m_0) \nonumber \\
&=& 1 + \int_{m_0}^m dm_1 \int d^4x_1 S(x_0 - x_1, m_1) \nonumber \\
& & \widehat{{\cal M}}(x_1, m_1, m_0).
\label{eq16}
\end{eqnarray}
Iteration of (\ref{eq16}) yields the perturbation solution

\begin{eqnarray}
& &\widehat{{\cal M}}(x_0, m, m_0) = 
1 + \int_{m_0}^m dm_1 \int d^4x_1 \, S(x_0 - x_1, m_1) \nonumber \\
&+& \int_{m_0}^m dm_1 \int d^4x_1 \, S(x_0 - x_1, m_1)
\int_{m_0}^{m_1} dm_2 \int d^4x_2 \nonumber \\
& & S(x_1 - x_2, m_2) \; + \; \cdots \nonumber \\
&+& \int_{m_0}^m dm_1 \int d^4x_1 \, S(x_0 - x_1, m_1)
\int_{m_0}^{m_1} dm_2 \int d^4x_2 \nonumber \\
& & S(x_1 - x_2, m_2) \; \cdots \; 
\int_{m_0}^{m_{n-1}} dm_n \int d^4x_n \nonumber \\
& & S(x_{n-1} - x_n, m_n) \; + \; \cdots
\label{eq17}
\end{eqnarray}

It is clear from the integration sequences in (\ref{eq17}) that $m_1
\geq m_2 \geq \; \cdots \; \geq m_{n-1} \geq m_n$. Therefore, each
product of Dirac propagators $S$ in (\ref{eq17}) is in mass ordered
form (which is analogous to the time ordering of the various products
of Hamiltonian operators in the time evolution operator
$\widehat{U}(t,t_0)$ \cite{bjo}). We now define a mass ordering
operator $\widehat{M}$ for the product of two quantities $A_1(m_1)$ and
$A_2(m_2)$ by

\begin{equation}
\widehat{M}[A_1(m_1) A_2(m_2)] = \left\{
\begin{array}{ll}
A_1(m_1) A_2(m_2) & \mbox{if $m_1 \geq m_2$} \\
A_2(m_2) A_1(m_1) & \mbox{if $m_1 < m_2$}.
\end{array} \right. 
\label{eq17.5}
\end{equation}
$\widehat{M}$ is defined in a similar manner for products of more than
two quantities $A_i(m_i)$. By using $\widehat{M}$ to retain the mass
ordering in (\ref{eq17}), we can now extend the mass integrations in
the various terms in (\ref{eq17}) to the full range from $m_0$ to $m$
by including an $\frac{1}{n !}$ compensation factor for the $n$--th
term. Thus,

\begin{eqnarray}
& &\widehat{{\cal M}}(x_0, m, m_0) = 
1 + \sum_{n=1}^\infty \frac{1}{n !}
\int_{m_0}^m dm_1 \int_{m_0}^m dm_2 \; \cdots \nonumber \\
& &\int_{m_0}^m dm_n \; \cdots \; 
\int d^4x_1 \int d^4x_2 \; \cdots \; \int d^4x_n \nonumber \\
& & \widehat{M}[S(x_0 - x_1, m_1) \; S(x_1 - x_2, m_2) \; \cdots \nonumber \\ 
& & S(x_{n-1} - x_n, m_n)]. 
\label{eq18}
\end{eqnarray}
Since (\ref{eq18}) has the form of a Taylor series expansion of the
exponential function, we write $\widehat{{\cal M}}(x, m, m_0)$ on the
shorthand form

\begin{equation}
\widehat{{\cal M}}(x, m, m_0) = \widehat{M}
\exp \Biggl(\int_{m_0}^m dm' \int d^4x' S(x - x', m') \Biggr). 
\label{eq19}
\end{equation}
It is seen that the above derivation of the expression for
$\widehat{{\cal M}}(x, m, m_0)$ is similar to the derivation
of the expression for the standard time evolution
operator $\widehat{U}(t,t_0)$ \cite{bjo,sak}. 

For definiteness, we represent the Feynman propagator by the
four--dimensional Fourier space integral solution of (\ref{eq12.5})
\cite{ryd},

\begin{equation}
\Delta_F(x, m) = \frac{1}{(2 \pi)^4}
\int \frac{e^{-i k x}}{k^2 - m^2} \, d^4k,
\label{eq20}
\end{equation}
where $k$ is the four--dimensional wave vector 

\begin{equation}
k = k^\mu = (k_0, \vec{k}).
\label{eq21}
\end{equation}
For simplicity of notation, we here assume that the $k$--integration
involves the proper path around the two poles on the $k_0$ axis [thus
omitting the usual infinitesimal parameter $\epsilon$ in the
denominator of (\ref{eq20})] \cite{ryd}. Substitution of (\ref{eq20})
into (\ref{eq12}) yields

\begin{equation}
S(x, m) = \frac{1}{(2 \pi)^4}
\int \frac{(\gamma^\mu k_\mu + m) e^{-i k x}}
{k^2 - m^2} \, d^4k.
\label{eq22}
\end{equation}

One can now substitute (\ref{eq22}) into (\ref{eq18}) and carry out the
mass integrations at the various orders in (\ref{eq18}). To first order
we have

\begin{eqnarray}
& &\widehat{{\cal M}}^1(x_0, m, m_0) = 
\int_{m_0}^m dm_1 \int d^4x_1 S(x_0 - x_1, m_1) \nonumber \\
&=&\frac{1}{(2 \pi)^4}
\int_{m_0}^m dm_1 \int d^4x_1 \int d^4k 
\frac{(\gamma^\mu k_\mu + m_1) e^{-i k (x_0 - x_1)}}
{k^2 - m_1^2} \nonumber \\
&=& \frac{1}{(2 \pi)^4} \int d^4k \int d^4x_1 
\Biggl[\frac{\gamma^\mu k_\mu}{|k|} 
\Biggl(\tanh^{-1}\frac{m}{|k|} \nonumber \\
&-& \tanh^{-1}\frac{m_0}{|k|}\Biggr)
+ \frac{1}{2} \ln\frac{m_0^2 - k^2}{m^2 - k^2}\Biggr]
e^{-i k (x_0 - x_1)},
\label{eq23}
\end{eqnarray}
where $|k| = \sqrt{k_0^2 - \vec{k}^2}$. 

In accord with the previous discussion, we view the propagation of the
attributive field $\widetilde{\psi}(x_0, m_0)$ in (\ref{eq10}) as
taking place through both a spacetime channel and a mass channel. To
propagate $\widetilde{\psi}(x_0, m_0)$ to $\widetilde{\psi}(x, m)$, one
may propagate the field through spacetime only by use of an extended
Dirac propagator $\widetilde{S}(x, m, m_0)$ that contains in it the
mass channel propagation from $\widetilde{\psi}(x_0, m_0)$ to
$\widetilde{\psi}(x_0, m)$. From (\ref{eq23}), we see that the
first-order term of a perturbation expansion of $\widetilde{S}(x, m,
m_0)$ has the Fourier space representation

\begin{eqnarray}
& &\widetilde{S}^1(x, m, m_0) \nonumber \\
&=& \frac{1}{(2 \pi)^4} \int  
\Biggl[ \frac{\gamma^\mu k_\mu}{|k|} 
\Biggl(\tanh^{-1}\frac{m}{|k|} 
- \tanh^{-1}\frac{m_0}{|k|}\Biggr) \nonumber \\
&+& \frac{1}{2} \ln\frac{m_0^2 - k^2}{m^2 - k^2} \Biggr]
e^{-i k x} \, d^4k.
\label{eq24}
\end{eqnarray}
Higher order contributions to the extended propagator $\widetilde{S}(x,
m, m_0)$ can readily be computed from (\ref{eq22}) and (\ref{eq18}).

A procedure similar to that used for the mass attribute of the extended
Dirac field $\widetilde{\psi}$ can be applied for the spin attribute of
$\widetilde{\psi}$. In analogy with the case above involving mass $m$,
the point of departure is to interpret the quantities $\gamma^\mu$ in
(\ref{eq8.5}) to represent the spin of $\widetilde{\psi}$. We use the
metric $g^{\mu\nu}$ of signature $(1, -1, -1, -1)$ to write
(\ref{eq8.5}) on the form

\begin{equation}
(i g^{\mu\lambda} \gamma_\lambda \partial_\mu - m)\psi(x) 
= I_{\text{int}}(x),
\label{eq25}
\end{equation}
and then differentiate (\ref{eq25}) with respect to $\gamma_\sigma$.
The result is [with the replacement of $\psi(x)$ by the extended
field $\widetilde{\psi}(x, \gamma^\sigma)$]

\begin{equation}
i g^{\mu\sigma} \partial_\mu \widetilde{\psi}(x, \gamma^\sigma)
+ i \gamma^\mu \partial_\mu 
\frac{\partial \widetilde{\psi}(x, \gamma^\sigma)}
{\partial \gamma_\sigma} 
- m \frac{\partial \widetilde{\psi}(x, \gamma^\sigma)}
{\partial \gamma_\sigma} = 0
\label{eq26}
\end{equation}
or
\begin{equation}
(i \gamma^\mu \partial_\mu - m)
\frac{\partial \widetilde{\psi}}{\partial \gamma_\sigma} 
+ i \partial^\sigma \widetilde{\psi} = 0.
\label{eq27}
\end{equation}

The resulting integro-differential equation [analogous to (\ref{eq12.6})]
now becomes

\begin{equation}
i \frac{\partial \widetilde{\psi}(x, \gamma^\sigma)}{\partial \gamma_\sigma} 
= \int S(x - x_1, m) \partial^\sigma \widetilde{\psi}(x_1, \gamma^\sigma) \, d^4x_1,
\label{eq28}
\end{equation}
which may be interpreted as describing an evolution of
$\widetilde{\psi}$ through a spin channel. Further analysis for each
component of $\gamma^\sigma$ can be carried out straightforwardly in a
manner analogous to the one above for $m$.

It is interesting to note that the attributive approach taken here,
which is based on the Dirac equation, leads to a specific relationship
between the spin-space and mass-space evolution of the extended Dirac
field. From (\ref{eq28}) and (\ref{eq12.6}) we see that these
non-spacetime evolutions are connected via the operational relation

\begin{equation}
i \frac{\partial}{\partial \gamma_\sigma} 
= \frac{\partial}{\partial m}  \frac{\partial}{\partial x_\sigma}
\label{eq29}
\end{equation}
between the spin ($\gamma^\sigma$), mass ($m$), and spacetime
($x^\sigma$) attributes of the extended Dirac field.

\section{Summary and Conclusions}

We have proposed that the spacetime location $x$ of particles
represented by a field $\widetilde{f}$ is nothing more than
an attribute of the field itself. This obviates the need for 
treating space and time in any special manner, as is done in
field theories of matter in particular, and physics in general.
As a point of departure compatible with this proposal, we 
have treated other attributes of the field in a way equivalent 
to the current treatment of spacetime in field theories of matter.
Specifically, we have suggested how field equations may be extended
when elevating non-spacetime attributes of a field to the same status
as that of the spacetime attribute. We finally applied this to the 
case of an attributive extension of the Dirac equation.

\acknowledgements

This research was supported by NSF Grant No. PHY94-15583.

\end{document}